\magnification \magstep 1
\centerline {\bf SYSTEMS AND SUBSYSTEMS }
\centerline {\bf IN QUANTUM COMMUNICATION}

\vskip 1 true cm
\centerline {\bf V.V. Belokurov, O.A. Khrustalev, V.A. Sadovnichy}
\centerline {\bf  and O.D. Timofeevskaya}

\centerline {M.V. Lomonosov Moscow State University, Moscow, Russia}
\centerline {{\it Email:} khrust@goa.bog.msu.ru}
\vskip 1 true cm

\leftskip -7 mm 
\rightskip 7 mm
\centerline {\it  In this paper we develop the Conditional
 Density Matrix formalism}
\centerline {\it for adequate description of division and unification of quantum
 systems.}
\centerline {\it   Applications of this approach to the
 descriptions of parapositronium,}
\centerline {\it  quantum teleportation and
 others examples are discussed.}
\vskip 3 true cm

\centerline {\bf INTRODUCTION}
\vskip 0.5 true cm

Recent progress in quantum communications has caused the great
interest to the problems connected with divisions of quantum
systems into subsystems and reunifications of subsystems into a
joint system.

Although general theory of such processes was proposed in 1927
{\bf [von Neumann 1927]}, so far, a division of a quantum system
into subsystems is usually described in a fictitious
manner.  As an example, here we quote  the classical paper
on the photon teleportation {\bf[Bouwmeester et al 1997]}. 
Describing  the photon teleportation experiment they write:

{\it The entangled state
contains no information on the individual particles; it only
indicates that two particles will be in the opposite states. The
important property of an entangled pair that as soon as a
measurement on one particles projects it, say, onto
$|\leftrightarrow>$ the state of the other one is determined to be
$|\updownarrow>$, and vice versa. How could a measurement on one
of the particles instantaneously influence the state of the other
particle, which can be arbitrary far away?  Einstein, among many
other distinguished physcists, could simply not accept this
"spooky action at a distance". But this property of entangled
states has been demonstrated by numerous experiments.

}

Nevertheless Einstein was quite right in his non-acceptance of such
point of view. In this paper we develope the correct approach to 
describe the phenomena completely adequate to the physical problem.
The basic notion of our approach is Conditional Density Matrix.

\vskip 1 true cm

\centerline {\bf CONDITIONAL DENSITY MATRIX}

\vskip 0.5 true cm

 Consider two systems $S_{1}$ and $S_{2}$. The
joint system is denoted as $S_{12}$.

The principal question that we want to answer here is how the
states of the subsystems  are related to the
state of the joint system, and vice versa.

Let ${\rho}_{1}$ and ${\rho}_{2}$ be the density matrices of the
systems $S_{1}$ and $S_{2}$.

If at least one of the states ${\rho}_{1}$ or ${\rho}_{2}$ is pure
(i.e. ${{\rho}_{i}}^{2} = {\rho}_{i}$) then these states determine
the state of the compound system $S_{12}$ uniquely:
$$   \rho \quad = \quad {\rho}_{1} \otimes {\rho}_{2}
$$

If the state of the system $S_{12}$ is ${\rho}_{12}$ then the
state of the system $S_{1}$ is determined by the following
equation:
$$   {\rho}_{1} \quad = \quad Tr_{2}\ ({\rho}_{12}).
$$

Now we can define the conditional density matrix.

 If the state of
the system $S_{12}$ is ${\rho}_{12}$ then the state of the system
$S_{1}$ upon the condition that the system $S_{2}$ is in the pure
state $   {\rho}_{2} ,\ {{\rho}_{2}}^{2} = {\rho}_{2} $ is
$$    {\rho}_{1/2} \quad = \quad
        {Tr_{2}\ ({\rho}_{2} \ {\rho}_{12})
      \over Tr \ ({{\rho}_{2}}\ {\rho}_{12})}.
$$

\vskip 1 true cm

\centerline {\bf Example: Parapositronium}
\vskip 0.5 true cm

As an example we consider parapositronium - the system consisting
of an electron and a positron. The total spin of the system is
equal to zero. In this case the nonrelativistic approximation is
valid and the state vector of the system is represented in the
form of the product
$$  {\Psi}({\vec r}_{e},{\sigma}_{e}; {\vec r}_{p}, {\sigma}_{p})
      \quad = \quad
     {\Phi}({\vec r}_{e},{\vec r}_{p})
     \chi({\sigma}_{e},{\sigma}_{p}).
$$
The spin wave function is equal to
$$  \chi({\sigma}_{e},{\sigma}_{p})
      \quad = \quad {1 \over \sqrt{2}}
   ({\chi}_{\vec n}({\sigma}_{e}){\chi}_{-{\vec n}}({\sigma}_{p})
         \quad  - \quad
  {\chi}_{\vec n}({\sigma}_{p}){\chi}_{-{\vec n}}({\sigma}_{e})).
$$
Here ${\chi}_{\vec n}(\sigma)$ and ${\chi}_{(-{\vec n})}(\sigma)$
are the eigenvectors of the operator that projects spin onto the
vector $\vec n$:
$$  ({\vec {\sigma}}{\vec n})\ {\chi}_{\vec n}(\sigma)
        \quad = \quad
       {\chi}_{\vec n}(\sigma),
$$
$$  ({\vec {\sigma}}{\vec n})\ {\chi}_{(-{\vec n})}(\sigma)
        \quad = \quad - \
       {\chi}_{(-{\vec n})}(\sigma).
$$
The spin density matrix of the system is determined by the
operator with the kernel
$$  \rho({\sigma};{\sigma}^{'})
     \quad = \quad
    {\chi}({\sigma}_{e},{\sigma}_{p})\
    {{\chi}}^{*}({\sigma}^{'}_{e},{\sigma}^{'}_{p}),
$$
The spin density matrix of the electron is
$$  {\rho}_{e}({\sigma},{\sigma}^{'})
     \quad = \quad
  \sum_{\xi}\ {\chi}({\sigma},\xi)\
    {{\chi}}^{*}({\sigma}^{'}, \xi)
     \quad = \quad
$$
$$       {1 \over 2}\
  ({\chi}_{\vec n}(\sigma)\ {\chi}_{(-{\vec n})}({\sigma}^{'})
         \  + \
  {\chi}_{\vec n}(\sigma)\ {\chi}_{(-{\vec n})}({\sigma}^{'}))
      \quad = \quad {1 \over 2}E(\sigma,{\sigma}^{'}).
$$
In this state the electron is completely unpolarized.

If an electron passes through polarization filter then
 the pass probability is independent of
the filter orientation. The same fact is valid for the positron if
its spin state is measured independently of the electron.

Now let us consider quite different experiment. Namely,  the
positron passes through the polarization filter and  the electron
polarization is simultaneously measured. The operator that
projects the positron spin onto the vector $\vec m$ (determined by
the filter) is given by the kernel
$$  P(\sigma,{\sigma}^{'})  \quad = \quad
  {\chi}_{\vec m}(\sigma)\ {\chi}^{*}_{{\vec m}}({\sigma}^{'}).
$$
Now the conditional density matrix of the electron equals to
$$   {\rho}_{e/p}(\sigma,{\sigma}^{'})
        \ = \
  {\sum_{(\sigma,{\sigma}^{'})}
   {\chi}_{\vec m}(\sigma)\ {\chi}^{*}_{\vec m}({\sigma}^{'})\
       {\chi}({\sigma}_{e},{\sigma}^{'})\
      {{\chi}^{*}}({\sigma}^{'}_{e},\sigma)
      \over
  \sum_{(\xi,\sigma,{\sigma}^{'})}
   {\chi}_{\vec m}(\sigma)\ {\chi}^{*}_{\vec m}({\sigma}^{'})\
       {\chi}(\xi,{\sigma}^{'})\
      {{\chi}^{*}}(\xi,\sigma)}.
$$
The result of the summation is
$$   {\rho}_{e/p}(\sigma,{\sigma}^{'})
         \quad = \quad
   {\chi}_{(-\vec m )}(\sigma)\ {\chi}^{*}_{(-\vec m )}({\sigma}^{'}).
$$

Thus, if the polarization of the positron is well defined
then the electron appears to be polarized in the opposite direction.

\vskip 1 true cm

\centerline {\bf TELEPORTATION}
\vskip 0.5 true cm

In the Innsbruck experiment on a photon state teleportation, the
initial state of the system is the result of the unification of
the pair of photons 1 and 2 being in the antisymmetric state
${\chi}({\sigma}_{1},{\sigma}_{2})$ with summary angular momentum
equal to zero and the photon 3 being in the state ${\chi}_{\vec
m}({\sigma}_{3})$ (that is, being polarized along the vector $\vec
m $). The joint system state is given by the density matrix
$$   \rho(\sigma, {\sigma}^{'})
       \quad = \quad
       {\Psi}(\sigma){{\Psi}^{*}}({\sigma}^{'}),
$$
where the wave function of the joint system is the product
$$   {\Psi}(\sigma)
     \quad = \quad
     {\chi}({\sigma}_{1},{\sigma}_{2})\ {\chi}_{\vec m}({\sigma}_{3}).
$$
Considering then the photon 2 only (without fixing the states of
the photons 1 and 3) we find the photon 2 to be completely
unpolarized with the density matrix
$$  {\rho}({\sigma}_{2},{\sigma}_{2}^{'})
       \  = \
    Tr_{(1,3)}\
{\rho}({\sigma}_{1},{\sigma}_{2},{\sigma}_{3};
{\sigma}_{1},{\sigma}_{2}^{'},{\sigma}_{3})
    \  = \
   {1 \over 2}\ E({\sigma}_{2},{\sigma}_{2}^{'}).
$$
However, if the photon 2 is registered when the state of the
photons 1 and 3 has been determined to be
${\chi}({\sigma}_{1},{\sigma}_{3})$ then the state of the photon 2
is given by the conditional density matrix
$$ {\rho}_{2/\lbrace 1,3 \rbrace}
      \quad = \quad
   {Tr_{(1,3)}\ (P_{1,3}\ {\rho}_{1,2,3}) \over
      Tr\ (P_{1,3}\ {\rho}_{1,2,3})}.
$$
Here $ P_{1,3}$ is the projection operator
$$   P_{1,3} \quad = \quad
      {\chi}({\sigma}_{1},{\sigma}_{3})\ {\chi}^{*}({\sigma}_{1},{\sigma}_{3}).
$$
To evaluate the conditional density matrix it is convenient to
preliminary find the vectors
$$   {\phi}({\sigma}_{1})  \quad = \quad
       \sum_{3}\ {\chi}^{*}_{\vec m}({\sigma}_{3})\ {\chi}({\sigma}_{1},{\sigma}_{3})
$$
and
$$  {\theta}({\sigma}_{2}) \quad = \quad
       \sum_{1}\
       {{\phi}^{*}}({\sigma}_{1})\ {\chi}({\sigma}_{1},{\sigma}_{2}).
$$
The vector $\theta$ equals to
$$  {\theta}({\sigma}_{2}) \quad = \quad
      - {1 \over 2}\ {\chi}_{\vec m}({\sigma}_{2})
$$
and the conditional density matrix of the photon 2 appears to be
equal to
$$ {\rho}_{2/\lbrace 1,3 \rbrace}
      \quad = \quad
      {\chi}_{\vec m}({\sigma}_{2})\ {{\chi}}^{*}_{\vec m}({\sigma}_{2}^{'}).
$$
Thus, if the subsystem consisting of the photons 1 and 3 is forced
to be in the antisymmetric state
${\chi}({\sigma}_{1},{\sigma}_{3})$ (with total angular momentum
equal to zero) then the photon 2 appears to be polarized along the
vector ${\vec m}$.

\vskip 1 true cm

\centerline {\bf PAIRS OF POLARIZED PHOTONS}
\vskip 0.5 true cm

Now consider  a modification of the Innsbruck experiment. Let
there be two pairs of photons $(1,\ 2)$ and $(3,\ 4)$. Suppose
that each pair is in the pure antisymmetric state $\chi$. The
spin part of the density matrix of the total system is given
by the equation
$$  {\rho}(\sigma,{\sigma}^{'})
       \quad = \quad
     {\Psi}(\sigma)\ {{\Psi}^{*}}({\sigma}^{'}).
$$
The wave function of the total system is the product of the wave
functions of the subsystems
$$  {\Psi}(\sigma)
    \quad = \quad
     {\chi}({\sigma}_{1},{\sigma}_{2})\ {\chi}({\sigma}_{3},{\sigma}_{4}).
$$
If the photons 2 and 4 are polarised along
 ${\chi}_{\vec m}({\sigma}_{2})$ and ${\chi}_{\vec
s}({\sigma}_{4})$ 
  then the wave function of the system is
transformed into
$$  {\Phi}(\sigma)
      \quad = \quad
 {\chi}_{\vec n}({\sigma}_{1})\ {\chi}_{\vec m}({\sigma}_{2})
\   {\chi}_{\vec r}({\sigma}_{3})\ {\chi}_{\vec s}({\sigma}_{4}).
$$
Here ${\vec n},\  {\vec m}$ and ${\vec r},\  {\vec s}$ are pairs
of mutually orthogonal vectors.

Now the conditional density matrix of the pair of photons 1 and 3
is
$$  {\rho}_{(1,3)/(2,4)}(\sigma,{\sigma}^{'})
     \quad = \quad
        {\Psi}({\sigma}_{1},{\sigma}_{3})\
        {{\Psi}^{*}}({\sigma}_{1}^{'},{\sigma}_{3}^{'}).
$$
The wave function of the pair is the product of wave
functions of each photon with definite polarization
$$   {\Psi}({\sigma}_{1},{\sigma}_{3})
        \quad = \quad
   {\chi}_{\vec n}({\sigma}_{1})\ {\chi}_{\vec r}({\sigma}_{3}) .
$$

Pairs of polarized photons appeart to be very useful in quantum 
communication.

\vskip 1 true cm

\centerline {\bf QUANTUM REALIZATION }
 \centerline {\bf  OF VERNAM
COMMUNICATION SCHEME}
\vskip 0.5 true cm

Let us recall the main idea of Vernam communication scheme {\bf
[Vernam 1926]}. In this scheme, Alice encrypts her message (a
string of bits denoted by the binary number $m_{1}$)
 using a randomly generated key $k$.
She simply adds each bit of the message with the corresponding bit
of the key to obtain the scrambled text ($ s = m_{1} \oplus k $,
where $\oplus$ denotes the binary addition modulo 2 without
carry). It is then sent to Bob, who decrypts the message by
subtracting the key  ($s \ominus k = m_{1} \oplus k \ominus k =
m_{1}$). Because the bits of the scrambled text are as random as
those of the key, they do not contain any information. This
cryptosystem is thus provable secure in sense of information
theory. Actually, today this is the only provably secure
cryptosystem!

Although perfectly secure, the problem with this security is that
it is essential that Alice and Bob possess a common secret key,
which must be at least as long as the message itself. They can
only use the key for a single encryption. If they used the key
more than once, Eve could record all of the scrambled messages and
start to build up a picture of the plain texts and thus also of
the key. (If Eve recorded two different messages encrypted with
the same key, she could add the scrambled text to obtain the sum
of the plain texts: $s_{1} \oplus s_{2} = m_{1} \oplus k \oplus
m_{2} \oplus k = m_{1} \oplus m_{2} \oplus k \oplus k = m_{1}
\oplus m_{2}$, where we used the fact that $\oplus$ is
commutative.) Furthermore, the key has to be transmitted by some
trusted means, such as a courier, or through a personal meeting
between Alice and Bob. This procedure may be complex and
expensive, and even may lead to a loophole in the system.

With the help of pairs of polarized photons we can overcome the
shortcomings of the classical realization of Vernam scheme.
Suppose Alice sends to Bob pairs of polarized photons obtained
according to the rules described in the previous section. Note
that the concrete photons' polarizations are set up in Alice's
laboratory and Eve does not know them. If the polarization of the
photon 1 is set up by a random binary number $p_{i}$ and the
polarization of the photon 3 is set up by a number $m_{i} \oplus
p_{i}$ then each photon (when considered separately) does not
carry any information. However, Bob after obtaining these photons
can add corresponding binary numbers and get the number $m_{i}$
containing the information ($m_{i} \oplus p_{i} \oplus
p_{i}=m_{i}$).

In this scheme, a secret code is created during the process of
sending and is transferred to Bob together with the information.
It makes the usage of the scheme completely secure.

\vskip 1 true cm

\centerline {\bf REFERENCES}
\vskip 0.5 true cm

{\bf von Neumann 1927} --- J. von Neumann,  G\"ott. Nach. pp.
1--57, 245 -- 272, {\bf 1927}. See, also, J. von Neumann,
Mathematische Grundlagen der Quantenmechanik, Berlin, 1932.

{\bf Bouwmeester et al 1997} --- D. Bouwmeester, J.-W. Pan, K.
Mattle, M. Eibl, H. Weinfurter, A. Zeillinger, Nature, {\bf 390},
pp. 577--579, {\bf 1997}.

{\bf Vernam 1926} --- G. Vernam, Cipher printing telegraph systems
for secret wire and radio telegraph communications, J. Am. Inst.
of Electrical Engineers, {\bf 45}, pp. 109-115, {\bf 1926}.

\bye